\documentclass[11pt]{article}
\usepackage{moriond,epsfig}

\def\Journal#1#2#3#4{{#1} {\bf #2}, #3 (#4)}

\def\Nat{{\em Nature}}
\def\Sci{{\em Science}}

\def\AA{{\em Aston. \& Astrophys.}}
\def\ApJ{{\em Astrophys. J.}}
\def\PR{{\em Phys. Rep.}}

\def\MNRAS{{\em Mon. Not. R. Astron. Soc.}}
\def\ANYAS{{\em Ann. NY Acad. Sci.}}
\def\NJP{{\em New J. Phys.}}

\def\be{\begin{equation}}
\def\ee{\end{equation}}
\def\bea{\begin{eqnarray}}
\def\eea{\end{eqnarray}}

\newcommand{\lesssim}{\lower.5ex\hbox{$\; \buildrel < \over\sim \;$}}
\newcommand{\gtrsim}{\lower.5ex\hbox{$\; \buildrel > \over\sim \;$}}

\begin{document}
\vspace*{4cm}
\title{GRB Theory in the Fermi Era}

\author{ J. Granot}

\address{{\rm on behalf of the Fermi LAT and GBM collaborations}\\ \vspace{0.1cm}
Centre for Astrophysics Research, 
University of Hertfordshire,
College Lane, Hatfield AL10 9AB, UK}

\maketitle\abstracts{
Before the launch of the Fermi Gamma-ray Space Telescope there were
only a handful of gamma-ray bursts (GRBs) detected at high energies
(above $100\;$MeV), while several different suggestions have been made
for possible high-energy emission sites and mechanisms. Here I briefly
review some of the theoretical expectations for high-energy emission
from GRBs, outline some of the hopes for improving our understanding
of GRB physics through Fermi observations of the prompt GRB emission
or the early afterglow (first few hours after the GRB), and summarize
what we have learned so far from the existing Fermi GRB observations
(over its first half-year of operation). Highlights include the first
detection of $>\;$GeV emission from a short GRB, as well as detailed
temporal and spectral information for the first GRB with $>\;$GeV
emission and a measured redshift, that has the highest measured
apparent (isotropic equivalent) radiated energy output (for any GRB),
the largest lower limit on the bulk Lorentz factor of the emitting
region, and constrains possible Lorentz invariance violation by
placing a robust lower limit on the quantum gravity mass.}

\section{Introduction: pre-Fermi high-energy GRB observations}

High-energy emission from GRBs was first detected by the Energetic
Gamma-Ray Experiment Telescope (EGRET) on-board the Compton Gamma-Ray
Observatory (CGRO; 1991$-$2000). While EGRET detected only 
five GRBs with its Spark Chambers ($20\;$MeV -- $30\;$GeV) and a few GRBs
with its Total Absorption Shower Counter (TASC; $1-200\;$MeV),
these events already showed diversity. Most noteworthy are GRB~940217,
with high-energy emission lasting up to $\sim 1.5\;$hr after the GRB
including an $18\;$GeV photon after $\sim 1.3\;$hr,~\cite{Hurley93}
and GRB~941017 which had a distinct high-energy spectral
component~\cite{Gonzalez03} detected up to $\sim 200\;$MeV with $\nu
F_\nu \propto \nu$. This high-energy spectral component had $\sim 3$
times more energy and lasted longer ($\sim 200\;$s) than the
low-energy (hard X-ray to soft gamma-ray) spectral component (which
lasted several tens of seconds), and may be naturally explained as
inverse-Compton emission from the forward-reverse shock system that is
formed as the ultra-relativistic GRB outflow is decelerated by the
external medium.~\cite{GrGu03,PW04} Nevertheless, better data are
needed in order to determine the origin of such high-energy spectral
components more conclusively. The Italian experiment Astro-rivelatore
Gamma a Immagini LEggero (AGILE; 2007$-$) has detected GRB~080514B at
energies up to $\sim 300\;$MeV, and the high-energy emission lasted
longer ($>13\;$s) than the low-energy emission ($\sim
7\;$s).~\cite{Giuliani08}

Fermi has raised great expectations for probing the high-energy
emission from GRBs as its Large Area Telescope (LAT; from $20\;$MeV to
$>\,$300$\;$GeV) significantly improves upon previous missions, mainly in
terms of its large effective area, small dead-time and large field of
view. Together with its Gamma-ray Burst Monitor (GBM; $8\;$keV --
$40\;$MeV) Fermi has an unprecedented energy range of $\sim 7.5$
decades, which is extremely useful for studying the GRB emission.

\section{Expectations from Fermi}

{\bf Prompt emission}:
most people hoped for, or even expected, the detection of a distinct
high-energy spectral component. Such a detection can shed light on the
prompt GRB emission mechanism at low energies (for which the $\nu
F_\nu$ spectrum which typically peaks at $E_{\rm peak}$ of around a
few hundred keV), which is still unclear, as well as the emission
mechanism at high energies. A high-energy spectral component may arise
either from leptonic processes, namely inverse-Compton scattering by
the same population of relativistic electrons responsible for the
observed low-energy prompt emission,~\cite{PM96,GuGr03} or from
hadronic processes~\cite{BD98} such as proton synchrotron, photo-pair
production, and pion production via photo-meson interaction or p-p
collisions, that may lead to pair cascades. Moreover, if the energy
output in such a high-energy spectral component is comparable to or
even larger than that in the low-energy spectral component (as seen by
EGRET for GRB~941017) then this will increase the already very tight
requirements on the source in terms of the total radiated energy and
the efficiency of the gamma-ray emission.~\cite{GKP06}

Many hopes were raised to detect a high-energy spectral cutoff
or steepening due to opacity to pair production ($\gamma\gamma \to
e^+e^-$) at the source.$\,$\cite{LS01,GCTdCeS08} Such a detection
would determine $\Gamma^{-2\beta}R$, where $\Gamma$ and $R$ are the
bulk Lorentz factor and distance from the source of the emitting
region, and $\beta$ is the (directly measurable) high-energy photon
index. Thus, it would determine both $\Gamma$ and $R$ for models (such
as the popular internal shocks model) in which $R
\sim \Gamma^2 c\Delta t$, where $\Delta t$ is the observed variability
time of the prompt GRB emission, or test whether this relation holds
if $\Gamma$ can be estimated independently (e.g. from the afterglow
onset time).

{\bf Longer lived high-energy emission}: several possible mechanisms
have been suggested for long lived high-energy emission from GRBs,
which may be detectable well after the end of the prompt GRB
emission. Synchrotron self-Compton (SSC -- the inverse-Compton
scattering of seed synchrotron photons emitted by the same population
of relativistic electrons) emission at GeV energies is expected from
the afterglow (i.e. the long lived forward shock going into the
external medium). Early on, when there is also a reverse shock going
into the ejecta, it can also produce inverse-Compton emission at
high energies, either via SSC, or by ``external-Compton'' (EC;
inverse-Compton scattering in which the seed photons are produced in a
different region), where reverse shock electrons scatter the
forward shock synchrotron photons, or vice versa.$\,$\cite{WDL01} In
some models$\,$\cite{SM00,GDM07} the reverse shock can be long-lived,
lasting for hours or even days, in which case such high-energy
emission involving the reverse shock would be similarly long-lived.
Other inverse-Compton processes involving two different emission
regions have also been suggested. In particular, the {\it Swift}
satellite detects flares in the early X-ray afterglow in about half of
the GRBs it observes, typically from hundreds to thousands of seconds
after the GRB. These X-ray flares are often attributed to sporadic
late-time activity of the central source, and are believed to be
emitted at a smaller radius than that of the contemporaneous afterglow
shock. In this scenario, EC may operate where afterglow electrons
scatter flare photons$\,$\cite{WLM06} or vice versa.$\,$\cite{Pan08}

Another mechanism that may produce long lived high-energy emission is
a pair echo. In this scenario $\gtrsim\;$TeV photons that escape the
source pair produce with the cosmic infrared background (or the cosmic
microwave background -- CMB), producing $e^+e^-$ pairs with $\sim {\rm
TeV}$ energies, that in turn inverse-Compton scatter CMB photons to
$\sim\;$GeV energies. This emission can potentially be detected up to
thousands of seconds after the GRB, if the inter-galactic magnetic
fields are sufficiently low ($\lesssim 10^{-20}\;$G for a correlation
length of $\sim 1\;$Mpc).~\cite{Plaga95,GuGr03} Finally, hadronic
processes involving high-energy cosmic-rays accelerated in the prompt
GRB emission region, or in the afterglow shock, could potentially
produce long-lived high-energy emission.

High-energy GRB observations by Fermi on a time scale of up to hours
after the GRB can either detect some of these emission components or
alternatively place interesting limits on them. In both cases, the
hope is that Fermi would thus be able to constrain the physical
conditions at the source and help determine the dominant high-energy
emission mechanisms.

\section{First results from Fermi}

{\bf GBM}: the GBM has a very wide field of view (full sky, half of which
is occulted by the Earth at any time) and is only slightly less
sensitive than the Burst and Transient Source Experiment (BATSE, that
was on-board the CGRO), thus resulting in a comparable (only slightly
lower) GRB detection rate of $\sim 250\;{\rm yr}^{-1}$, where $\sim
18\%$ of them are of the short duration spectrally hard class of
GRBs. A good fraction of GBM GRBs are within the LAT field of view.

{\bf LAT GRB detection rate}: during the first $\sim 9$ months of
operation Fermi LAT has clearly detected high-energy emission from 7
GRBs, corresponding to a detection rate of $\sim 9\;{\rm yr}^{-1}$. A
detailed comparison to the expected detection rate requires specifying
the number of detected photons above a certain energy. The preliminary
results (which suffer from a large statistical uncertainty due to the
small number of detected GRBs) are $\sim 7\,$--$\,8\;\;{\rm yr}^{-1}$
($\sim 1\,$--$\,2\;\;{\rm yr}^{-1}$) with at least 10 photons above
$100\;$MeV ($1\;$GeV). This is compatible (perhaps slightly lower but
well within the errors) with the expected rate~\cite{pre-launch} based
on a sample of bright BATSE GRBs for which the fit to a Band spectrum
over the BATSE energy range ($30\;$keV -- $2\;$MeV) is extrapolated
into the LAT energy range, and excluding cases with a rising $\nu
F_\nu$ spectrum at high energies (i.e. a high-energy photon index
$\beta > -2$).~\footnote{Such a hard high-energy photon index may be
an artifact of the limited energy range of the fit to BATSE data, and
even if such a hard spectrum is present in the BATSE range it is not
very likely that $\nu F_\nu$ continues to smoothly rise well into the
LAT energy range.} This suggests that, on average, there is no
significant excess (perhaps even a slight deficit) of high-energy
emission in the LAT energy range relative to such an extrapolation
from lower energies. Note that this expected detection rate (that is
close to the observed rate) is smaller than that for a larger sample
of BATSE bursts that includes events that are dimmer in the BATSE
range, some of which have $\beta \gtrsim -2$ and would be detectable
by the LAT upon extrapolation, thus increasing the expected LAT
detection rate. It should be noted, however, that for such GRBs that
are relatively dim in the BATSE range it is hard to determine the
value of $\beta$ very accurately, and it might suffer from some
systematic error.

{\bf GRB~081024B}: this GRB was detected by the LAT with more than 10
photons above $100\;$MeV, and is the first clearly short GRB that is
detected at high energies (up to a few GeV). Its spectrum is
consistent with a single Band function, similar to the LAT long
GRBs. Its high-energy emission ($>100\;$MeV) lasts about $3\;$s, while
its low-energy emission goes back to background levels after
$0.8\;$s. Even though it was not possible to determine its redshift
(due to the lack of an afterglow detection), the lack of a high-energy
cutoff in its spectrum up to the highest detected photon energies
implies a fairly high lower limit on its bulk Lorentz factor for any
reasonable redshift: $\Gamma_{\rm min}(z = 0.1) \approx 150$ while
$\Gamma_{\rm min}(z = 3)
\approx 900$. These values are significantly higher than the pre-Fermi 
conservative estimates for short GRBs~\cite{Nakar07}, that were based
on the prompt emission spectrum of many short BATSE GRBs being
well-fit by a power-law with a high-energy exponential cutoff, where
such an exponential cutoff at high energies results in a much lower
$\Gamma_{\rm min}$ compared to a (reasonably hard) power-law at high
energies.

\section{A minimal Lorentz factor of the emitting region
from compactness arguments}\label{Gamma_min}

The large isotropic equivalent luminosities ($L \sim
10^{50}-10^{53}\;{\rm erg\;s^{-1}}$) and short observed variability
time ($\Delta t \sim 1\;{\rm ms}-1\;$s) of GRBs would imply a huge
opacity to pair production ($\gamma\gamma \to e^+e^-$) within the
source ($\tau_{\gamma\gamma} \gg 1$) if the source (i.e. the emitting
region) is at rest or moving at a sub-relativistic velocity relative
to us. Neglecting cosmological factors of $(1+z)$ for simplicity, an
order of magnitude estimate of the optical depth at a dimensionless
photon energy $\varepsilon \equiv E_{\rm ph}/m_e c^2$ gives
$\tau_{\gamma\gamma} \sim
\sigma_T n_{\rm ph}(1/\varepsilon)R \sim
\sigma_T L_{1/\varepsilon}/(4\pi m_e c^3 R) \gtrsim 10^{14} 
(L_{1/\varepsilon}/10^{51}\;{\rm erg\;s^{-1}})(\Delta t/1\;{\rm
ms})^{-1}$, where $R \lesssim c\Delta t$ is the source size 
and $n_{\rm ph}(1/\varepsilon)$ is the number density of the target
photons (near the threshold for pair production) that provide most of
the opacity. Such a huge optical depth would result in a (quasi-)
thermal spectrum, in stark contrast with the significant high-energy
power-law tail observed in most GRBs. This is known as the compactness
problem.~\cite{Rud75} Its solution is that the source moves toward us
at a very high Lorentz factor, $\Gamma \gg 1$. This reduces
$\tau_{\gamma\gamma}$ due to three effects. First, the threshold for
pair production is $\varepsilon_1\varepsilon_2 >
2/(1-\cos\theta_{12})$ where $\varepsilon_1$ and $\varepsilon_2$ are
the two photon energies and $\theta_{12}$ is the angle between their
directions. For a source at rest, $\theta_{12} \sim 1$ and
$\varepsilon_1\varepsilon_2 \gtrsim 1$, while for a relativistic
source $\theta_{12} \sim 1/\Gamma$ (due to relativistic beaming) and
$\varepsilon_1\varepsilon_2 \gtrsim \Gamma^2$ (in the source rest
frame $\theta'_{12} \sim 1$ and $\varepsilon'_1\varepsilon'_2
\gtrsim 1$ where $\varepsilon' \sim \varepsilon/\Gamma$).
Thus $L_{1/\varepsilon}$ is replaced by $L_{\Gamma^2/\varepsilon} =
L_{1/\varepsilon}\Gamma^{2(1+\beta)}$, adding a factor of $\sim
\Gamma^{2(1+\beta)}$ to the expression for $\tau_{\gamma\gamma}$,
where $\beta$ is the high-energy photon index ($L_\varepsilon = L_0
\varepsilon^{1+\beta}$ in the relevant energy range). Second, $R$
should now represent the distance in the lab frame over which $n_{\rm
ph}$ is large enough to significantly contribute to
$\tau_{\gamma\gamma}$, i.e. roughly the distance of the emitting
region from the source, and $R \lesssim
\Gamma^2 c\Delta t$ is possible since $\Delta t \sim R/(c\Gamma^2)$ 
is the time delay in the arrival of photons from an angle of $\sim
1/\Gamma$ from the line of sight relative to the line of sight itself
for an emitting region with a radius of curvature $\sim R$ (in the lab
frame), as well as the difference in arrival time of two photons
emitted along the line of sight over a radial interval $\Delta R \sim
R$. This adds a factor of $\sim \Gamma^{-2}$ to the expression for
$\tau_{\gamma\gamma}$. Finally, there is a factor of
$1-\cos\theta_{12} \sim \Gamma^{-2}$ in the differential expression
for $\tau_{\gamma\gamma}$, due the the rate at which the photons pass
each other and have a chance of interacting (exactly parallel photons
will never interact). This results in an additional factor of $\sim
\Gamma^{-2}$ in the expression for $\tau_{\gamma\gamma}$. Altogether,
$\tau_{\gamma\gamma}$ includes a factor of $\sim \Gamma^{2(1-\beta)}$,
and since typically $-\beta \sim 2-3$, this typically requires $\Gamma
> \Gamma_{\rm min} \sim 100$ in order to achieve $\tau_{\gamma\gamma}
< 1$. In particular, when there is no high-energy cutoff or steepening
in the spectrum up to an observed photon energy of $\varepsilon_{\rm
max}$, then the requirement that $\tau_{\gamma\gamma}(\varepsilon_{\rm
max}) < 1$ leads to $\Gamma_{\rm min} \propto (L_0/\Delta
t)^{1/2(1-\beta)}
(\varepsilon_{\rm max})^{(-1-\beta)/2(1-\beta)}$.

\section{GRB~080916C}

GRB~080916C was the second GRB detected by the LAT and the brightest
so far. It had $>3000$ raw LAT counts (after background subtraction)
in the first $100\;$s, with 145 events above $100\;$MeV that could be
used for spectral analysis, and 14 photons above
$1\;$GeV.~\cite{Abdo09} The accurate localization by the LAT (to
within $\sim 0.1^\circ$) enabled the detection of its X-ray afterglow
after $17\;$hr in a follow-up observation by the {\it Swift} X-ray
telescope,~\cite{Swift08} which in turn provided a much better
localization (to within 1.9'') that enabled follow-up observations by
ground based telescopes and the detection of the optical/NIR afterglow
by the Gamma-Ray Burst Optical/Near-Infrared Detector (GROND), which
was able to measure a photometric redshift of $z = 4.35\pm
0.15$.~\cite{GROND09}

{\bf Energetics and beaming}:
GRB~080916C was a very bright long GRB. It had
a very high fluence of $f = 2.4 \times 10^{-4}\;{\rm erg\;cm^{-2}}$,
corresponding to an isotropic equivalent energy output of
$E_{\rm\gamma,iso} \approx 8.8 \times 10^{54}\;{\rm erg} \approx
4.9M_\odot c^2$, which is the highest measured so far for any GRB, and
strongly suggests that the outflow was collimated into a narrow jet,
in order to alleviate the otherwise very extreme energy requirements
from the source.

\begin{figure}
\psfig{figure=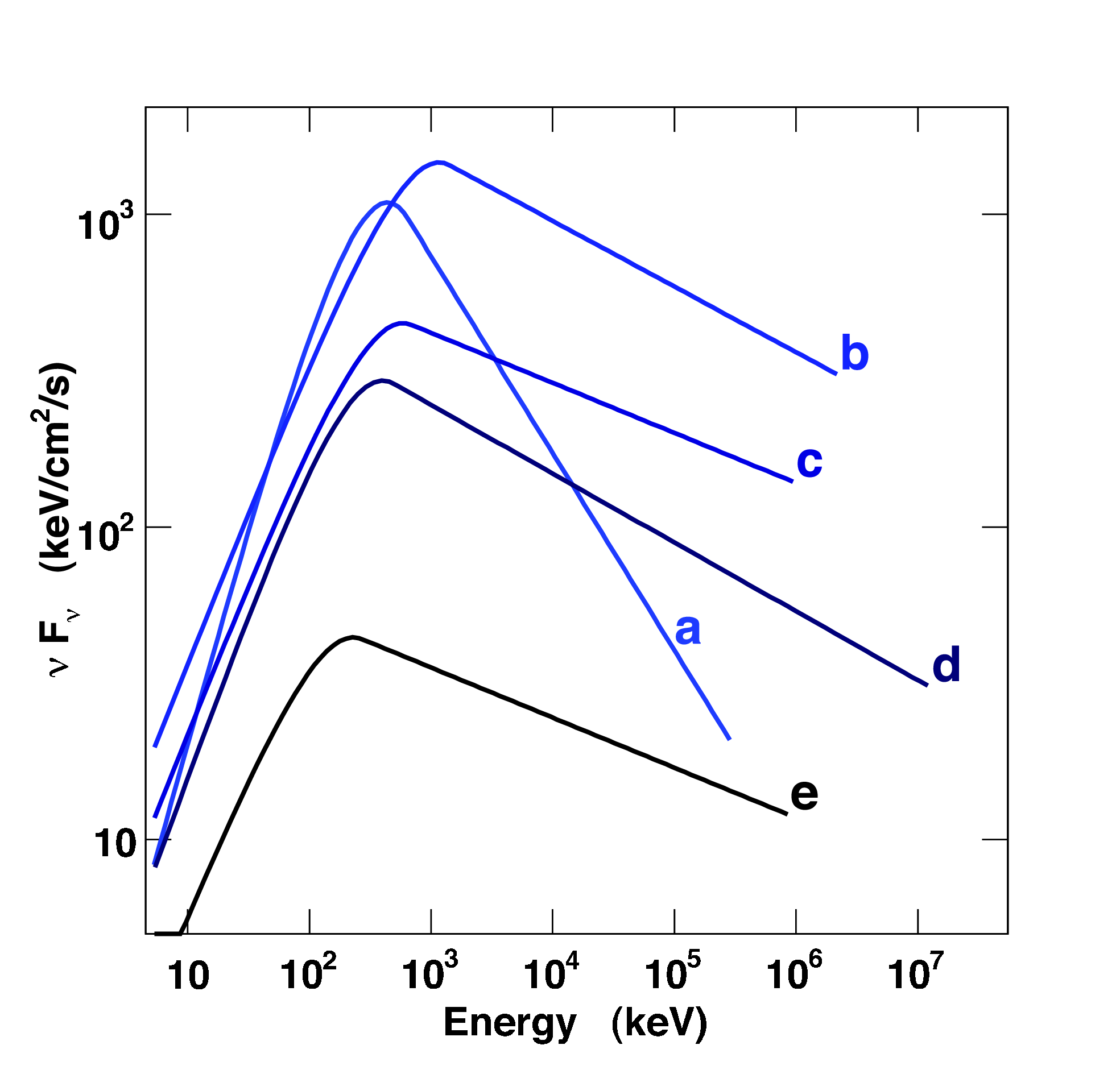,height=2.2in,width=2.0in}
\psfig{figure=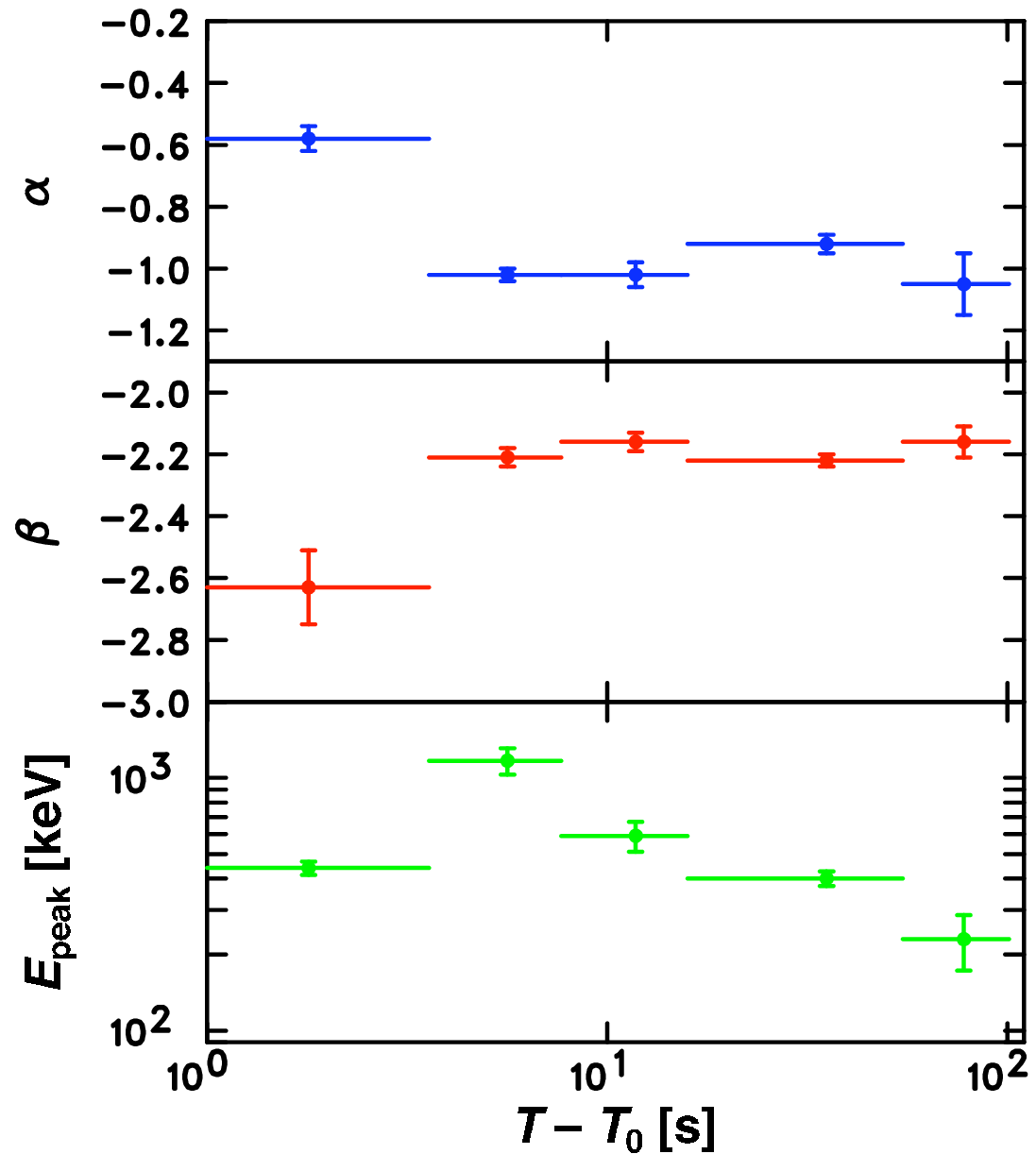,height=2.73in,width=2.0in}
\hspace{0.4cm}
\psfig{figure=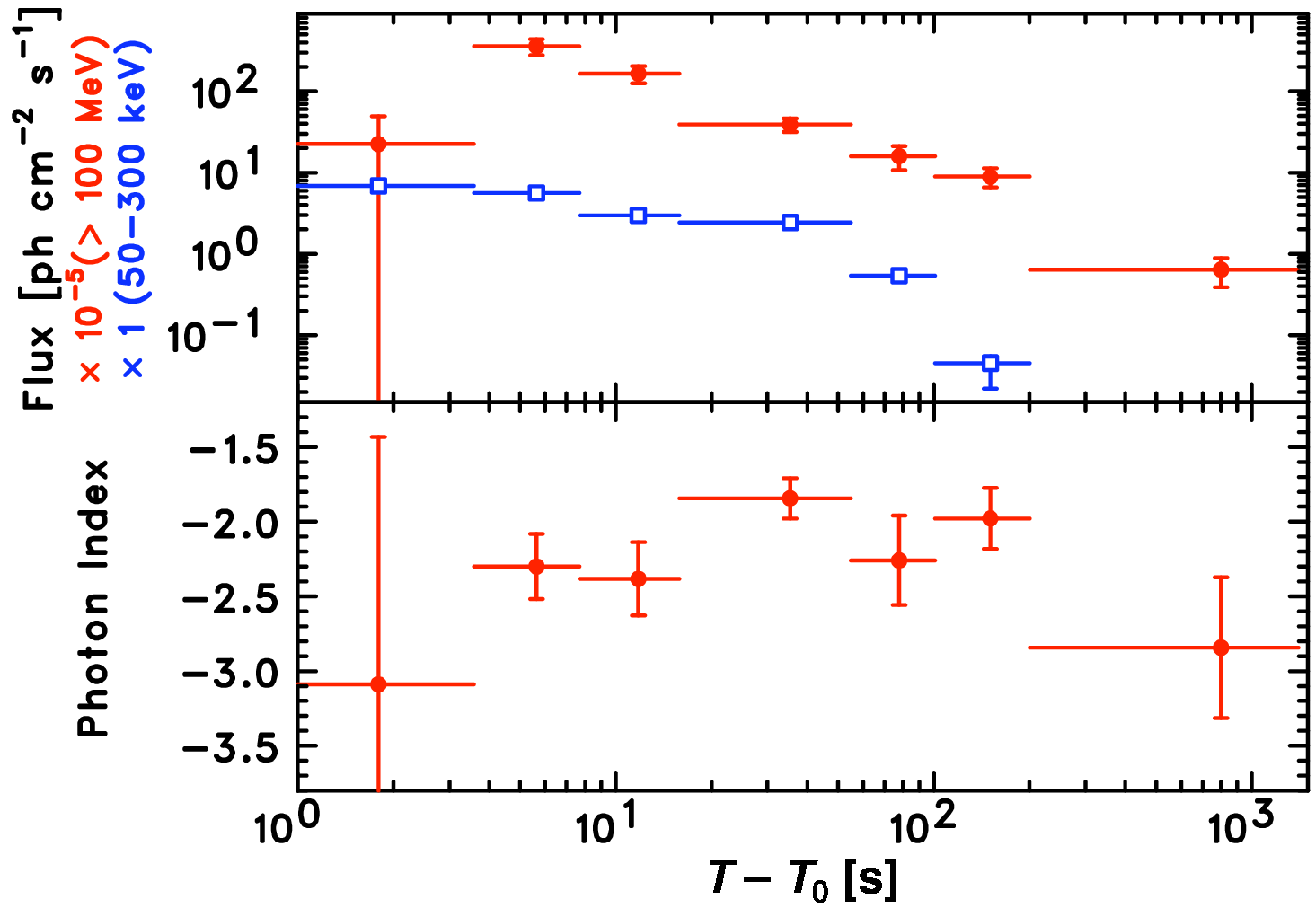,height=2.73in,width=2.0in}
\caption{\label{fig:spec}
Spectral evolution of GRB~080916C.~$^{21}\;$
{\bf Left panel}: The best-fit model $\nu F_\nu$ spectra for all five time
intervals.  The changing shapes show the evolution of the spectrum
over time. The curves end at the energy of the highest-energy photon
observed in each time interval. {\bf Middle panel}: Fit parameters for
the Band function -- the photon index at low ($\alpha$) and high
($\beta$) energies, and the photon energy ($E_{\rm peak}$) where the
$\nu F_\nu$ spectrum peaks -- as a function of time. Error bars
indicate $1\;\sigma$ uncertainty. {\bf Right panel}: Fluxes ({\it
top}) for the energy ranges $50-300\;$keV (blue open squares) and
$>100\;$MeV (red solid squares) and power-law index as a function of
the time from the GRB trigger time $T_0$ to $T_0 + 1400\;$s [({\it bottom})
LAT data only].\hfill\hfill}
\end{figure}

{\bf Spectral evolution}: the time resolved spectrum of the prompt
emission in GRB~080916C was analyzed in five different time bins
(chronologically labeled {\bf a}-{\bf e}; see {\it left} and {\it
middle panels} of Fig.~1) and found to be well-fit by a single Band
spectrum (featuring a smooth transition between two power-law
segments) in a combined fit of the LAT and GBM data.  The peak of the
$\nu F_\nu$ spectrum, $E_{\rm peak}$, first increases between the
first and second time bins, and then gradually decreases with time
(see {\it middle panel} of Fig.~1). The photon indices at low
energies, $\alpha$, and at high energies, $\beta$, both change between
the first and second time bins, $\alpha$ becoming softer and $\beta$
becoming harder, and are then consistent with remaining constant in
time.

{\bf Implications of a single dominant spectral component}: the fact
that the spectrum in time bins {\bf a}-{\bf e} is consistent with a
single Band function suggests that a single spectral component,
arising from a single emission mechanism, dominates throughout the
observed energy range, which cover 6 decades in energy (roughly
$10\;$keV -- $10\;$GeV).  This provides interesting constraints on any
emission mechanism.  For example, if the observed emission is
synchrotron radiation, then an SSC component may peak in the LAT
energy range, and the fact that it is not detected suggests that
either (i) it has a lower luminosity, its peak $\nu F_\nu$ being at
least $\sim 10$ times lower than that of the synchrotron component, if
the SSC component peak around several GeV, implying at least $\sim 10$
times more energy in relativistic electrons than in the magnetic field
in the emission region, or (ii) the SSC component may have a
comparable or even higher luminosity than that of the synchrotron
component if it peaks well above $10\;$GeV, in which case it will be
hard to detect it due to the smaller number of photons at higher
energies and attenuation due to pair production with the
extra-Galactic background light (EBL).

{\bf EBL}: in time bin {\bf d} there is weak evidence for a possible
high-energy excess relative to a Band spectrum.~\cite{Abdo09} The
chance probability of such an excess is $1\%$, and taking into account
the 5 trials (for bins {\bf a}-{\bf e}) it increases to $5\%$ (or
$2\;\sigma$). For some EBL models the optical depth for pair
production with the EBL of the highest energy detected photon,
$13.22^{+0.70}_{-1.54}\;$GeV, is $\tau_{\gamma\gamma} \sim
3\,$--$\,4$, in which case the significance of an additional
high-energy spectral component would be increased to $\sim
3\,$--$\,4\;\sigma$. Such a spectral component may increase the
already extreme apparent radiated energy in GRB~080916C. However, for
many other EBL models $\tau_{\gamma\gamma}(13\;{\rm GeV}) \ll 1$,
resulting in a mere $2\;\sigma$ hint of a possible excess, which is
not very significant.

{\bf Delayed high-energy onset}: the high-energy emission in
GRB~080916C starts $\sim 4\,$--$\,5\;$s after the low-energy emission. After
the onset of the LAT emission it quickly rises to a bright sharp peak
-- the main peak in the LAT lightcurve, which coincides with the
second peak in the GBM lightcurve (in time bin {\bf b}). If indeed the
observed spectrum in the GBM and LAT energy range is dominated by a
single spectral component, as suggested by the fact that it is
well-fit by a single Band spectrum, then the delayed HE onset may be
attributed mainly to a change in the high-energy photon index $\beta$
between the first and second pulses in the GBM lightcurve (as was
measured; see Fig. 1). This, in turn, may naturally occur if these two
pulses originated in two distinct physical regions (e.g. two sets of
colliding shells in the internal shocks model) with different physical
conditions, resulting in a different power-law index of the energy
distribution of the accelerated relativistic electron population that
is responsible for the observed emission.

Opacity effects do not work well as an alternative explanation since
there is no sign of a high-energy cutoff or steepening in the spectrum
(that must be present in the observed energy range in order for
opacity effects to be the major cause for the observed delayed onset).

Contribution from an additional spectral component at high energies
may be possible if together with the spectral component that dominates
at low energies the combined spectrum is still well-fit by a single
Band function (which is not always that easy to achieve). In this
case, however, it is not obvious why the effective value of $\beta$
(or the luminosity ratio of the two components) should remain constant
for the remainder of the GRB (time bins {\bf b}-{\bf e}). If the main
LAT peak is attributed to emission from the same physical region as
the first GBM peak (e.g. due to the gradual acceleration of
high-energy protons or heavier ions that produce pair cascades) then
it is not clear why it should coincide with the second GBM peak or why
the main LAT peak is as sharp as it is (as a much smoother peak would
be expected in this case).  Altogether, the exact cause for the
delayed high-energy onset is still not clear, and more detailed
modeling could help address this question.

{\bf Long lived high-energy emission}: while the low-energy emission
lasted several tens of seconds, with some low level emission detected
up to $200\;$s after the GRB trigger time, high-energy emission was
detected by the LAT for more than $1000\;$s. In particular, the LAT
detected emission above $100\;$MeV in two additional time bins (just
after time bins {\bf a}-{\bf e}), $100-200\;$s and $200-1400\;$s (see
{\it right panel} of Fig.~1). The $>100\;$MeV LAT flux decayed as
$t^{-1.2\pm 0.2}$ from several seconds and up to $1400\;$s, while
during the last time bin ($200-1400\;$s) the photon index was $\beta =
-2.8 \pm 0.5$. The GBM flux decayed more slowly ($\sim t^{-0.6}$) up
to $\sim 55\;$s, and faster ($\sim t^{-3.3}$) at later times (until
fading below detection threshold around $200\;$s).

Different possible mechanisms may account for such a long lived
high-energy emission. A natural possibility is afterglow SSC emission,
but spectral hardening is expected when this component becomes
dominant, and this is not seen in the data. Some time delay may be
caused by scattering of photons emitted at a smaller
radius~\cite{WLM06} (e.g. an inner set of colliding shells in the
internal shock model) or due pair cascades induced by
ultra-relativistic ions accelerated in the prompt emission
region.~\cite{DA06} In both cases, however, it might be hard to
produce the relatively slow decay rate, due to adiabatic losses on a
much shorter timescale (that of the observed prompt emission pulses).
Other options are scattering of photons from early X-ray flares
(undetected in this case, but detected by {\it Swift} in many other
GRBs) by afterglow electrons, or a pair echo. It is hard to
conclusively determine the exact mechanism at work here, but further
study may help distinguish between the different possibilities.

{\bf Comparison to other GRBs}: while there is a hint of a delayed
onset of the high-energy emission in other LAT GRBs, in those cases it
is not nearly as significant as in GRB~080916C. However, a longer
duration of the high-energy emission compared to the low-energy
emission appears in most LAT GRBs so far, and seems to be a common
feature in GRBs. Moreover, it also appeared in EGRET GRBs (especially
in GRB~940217) and in the AGILE GRB~080514B. It is hard to tell
whether the longer lived high-energy emission is from a similar
mechanism in all these cases or from different mechanisms in different
GRBs, due to the rather low photon statistics of this long-lived
emission and the lack of good broad-band monitoring of the
contemporaneous afterglow emission at lower frequencies (mainly X-ray
and optical). Nevertheless, such broad-band coverage may improve in
the near future and help in distinguishing between the different
possible physical origins of the long lasting high-energy emission.

{\bf Minimum Lorentz factor}: the very high isotropic equivalent
luminosity together with the fact that the spectrum did not show any
significant deviation from a Band spectrum up to the highest observed
photon energies (of $E_{\rm max} \gtrsim$ a few GeV) require a very
large bulk Lorentz factor of the emitting region, $\Gamma >
\Gamma_{\rm min}$, in order for the optical depth to pair production
in the source to satisfy $\tau_{\gamma\gamma}(E_{\rm max}) < 1$ (see
\S~\ref{Gamma_min}). For time bin {\bf d} this implies $\Gamma_{\rm
min} = 608 \pm 15$. For time bin {\bf b} $\Gamma_{\rm min} = 887 \pm
21$ for an observed variability time of $\Delta t = 2\;$s (the time
for a factor of $\sim 2$ GBM flux variation). A more careful
inspection of the low-energy lightcurve in time bin {\bf b} shows
significant variability at least down to timescales of $0.5\;$s, and
adopting $\Delta t = 0.5\;$s results in $\Gamma_{\rm min} \approx
1100$. Even the more conservative value of $\Gamma_{\rm min} \approx
900$ is more than twice the previous largest $\Gamma_{\rm min}$ for
any other GRB from opacity considerations.~\cite{LS01} Moreover, our
limit is more robust than previous ones, since in our case the target
photons that provide the opacity for the highest energy observed
photon are within the observed energy range ($E_{\rm ph} \ll E_{\rm
max}$), while for previous limits they were well above the observed
energy range ($E_{\rm ph} \gg E_{\rm max}$), and therefore it was not
clear whether they were indeed present at the source. Note that for
the conservative assumption that the photon spectrum reaches only up
to $E_{\rm max}$, $\Gamma_{\rm min} \lesssim (1+z)E_{\rm max}/m_ec^2
\approx 200(1+z)(E_{\rm max}/100\;{\rm MeV})$, and therefore a
large $\Gamma_{\rm min}$ requires the detection of high-energy
photons. Our lower limit on $\Gamma$ for time bin {\bf b} implies a
fairly large emission radius, $R \sim \Gamma^2 c\Delta t/(1+z) \gtrsim
10^{16}\;$cm.
 
{\bf Limits on Lorentz invariance violation}: some quantum gravity
models predict energy dispersion in the propagation speed of photons,
where high-energy photons travel slower~\footnote{In principle they
could also travel faster (or even faster in some photon energies and
slower in others). For GRB~080916C, however, there is no high-energy
photon detected before the onset of the low-energy emission (i.e. the
GRB trigger), and in fact the first of the 14 photons with energies
above $1\;$GeV arrives several seconds after the GRB
trigger. Therefore, a comparable or perhaps an even somewhat stricter
limit may be put on such a ``negative delay'' in the arrival time of
high-energy photons relative to low-energy photons.} than low-energy
photons.~\cite{LIV} The Lorentz invariance violating terms in the
dependence of the photon momentum $p_{\rm ph}$ on the photon energy
$E_{\rm ph}$ can be expressed as a power series,
\begin{equation}\label{eq:QG1}
\frac{p_{\rm ph}^2c^2}{E_{\rm ph}^2}-1 =
\sum_{k = 1}^\infty\left(\frac{E_{\rm ph}}{\xi_k M_{\rm Planck}c^2}\right)^k
= \sum_{k = 1}^\infty\left(\frac{E_{\rm ph}}{M_{{\rm QG},k}c^2}\right)^k\ ,
\end{equation}
in the ratio of $E_{\rm ph}$ and a typical energy scale $M_{\rm
QG,k}c^2 = \xi_k M_{\rm Planck}c^2$ for the $k^{\rm th}$ order, which
is expected to be of the order of the Planck scale, $M_{\rm planck} =
(\hbar c/G)^{1/2} \approx 1.22\times 10^{19}\;{\rm GeV/c^2}$. That is,
$\xi_k
\sim 1$ may naively be expected for the coefficients that are not infinite 
(some terms may not appear in this sum). Since we observe photons of
energy well below the Planck scale, the dominant Lorentz invariance
violating term is associated with the lowest order non-zero term in
the sum, of order $n$, which is usually assumed to be either first
order ($n = 1$) or second order ($n = 2$). The photon propagation
speed is given by the corresponding group velocity,
\begin{equation}
v_{\rm ph} = \frac{\partial E_{\rm ph}}{\partial p_{\rm ph}} \approx
c\left[1-\frac{n+1}{2}\left(\frac{E_{\rm ph}}{M_{{\rm QG},n}c^2}\right)^n\,\right]\ .
\end{equation}
Taking into account cosmological effects, this induces a time delay in
the arrival of a high-energy photon of energy $E_{\rm h}$, compared to
a low-energy photon of energy $E_{\rm l}$, of~\cite{JP08}
\begin{equation}\label{Dt_QG}
\Delta t \approx \frac{(1+n)}{2H_0}
\frac{(E_{\rm h}^n - E_{\rm l}^n)}{(M_{{\rm QG},n}c^2)^n}
\int_0^z\frac{(1+z')^n}{\sqrt{\Omega_M(1+z')^3+\Omega_\Lambda}}\,dz'\ .
\end{equation}
We apply this formula to the highest energy photon detected in
GRB~080916C, with an energy of $E_{\rm h} =
13.22^{+0.70}_{-1.54}\;$GeV, which arrived at $t = 16.54\;$s after the
GRB trigger (i.e. after the onset of the hard X-ray to soft gamma-ray,
sub-MeV emission: $E_l \sim 0.1\;$MeV). Since we have $E_{\rm
h}/E_{\rm l} \sim 10^5 \gg 1$, the term $E_{\rm h}^n$ in
eq.~(\ref{Dt_QG}) can be neglected, and $\Delta t \propto (E_{\rm
h}/M_{{\rm QG,n}})^n$. Since it is hard to associate the highest
energy photon with a particular spike in the low-energy lightcurve, we
make the conservative assumption that it was emitted sometime between
the GRB trigger and the time that it was observed, i.e. $\Delta t \leq
t$. This results in the following limits~\cite{Abdo09} for $n = 1$,
\begin{equation}
M_{{\rm QG},1} > (1.55\pm 0.04)\times 10^{18}
\left(\frac{E_{\rm h}}{13.22\;{\rm GeV}}\right)
\left(\frac{\Delta t}{16.54\;{\rm s}}\right)^{-1}\;{\rm GeV/c^2}\ ,
\end{equation}
and for $n = 2$, $M_{\rm QG,2} > (9.66\pm 0.22)\times 10^8(E_{\rm
h}/13.22\;{\rm GeV})(\Delta t/16.54\;{\rm s})^{-1/2}\;{\rm
GeV/c^2}$. Our limit for $n = 1$ is the strictest of its kind, and
only a factor of 10 below the Planck mass.

\section{Conclusions}

Fermi has raised great expectations that, similar to previous major
new relevant space missions, it would also significantly contribute to
the progress in the GRB field. The main expectations are to improve
our understanding of the prompt GRB emission mechanism and the
physical properties of the emission region, possibly by observing a
distinct high-energy spectral component or signatures of opacity to
pair production in the source, as well as improving our understanding
of the early afterglow. While most of these hopes will have to wait a
bit longer, Fermi has already provided some very interesting initial
results during its first half-year of operation. The spectrum of most
GRBs detected so far by both the GBM and the LAT is consistent with a
single Band function, suggestive of a single dominant emission
mechanism in the observed energy range, as is also suggested by the
LAT GRB detection rate.  Longer lived high-energy emission compared to
the low-energy emission (in some cases lasting $>10^3\;$s) appears to
be common in LAT GRBs.  Particularly interesting LAT GRBs are
GRB~081024B, the first clearly short GRB detected above $1\;$GeV, and
the exceptionally bright and energetic GRB~080916C that provided
a wealth of information leading to tight lower limits on the bulk
Lorentz factor of the emitting region and the quantum gravity
mass. Finally, there is still a lot to look forward to from Fermi.

\section*{Acknowledgments}
The author gratefully acknowledges a Royal Society Wolfson Research
Merit Award.

\end{document}